\documentclass{article}
\usepackage{graphicx}
\usepackage{amsmath}

\input{tcilatex}

\begin{document}

\title{\textbf{A random quantum key distribution by using \emph{Bell} states}%
}
\author{Chong Li\thanks{\texttt{\ lichong@student.dlut.edu}.cn}, He-Shan
Song, Ling Zhou \\
Department of Physics, Dalian University of Technology, \\
Dalian 116024, PR China \and Chun-Feng Wu \\
Department of Physics, Faculty of Science,\\
National University of Singapore, 119260, Singapore}
\maketitle

\begin{abstract}
We proposed a new scheme for quantum key distribution based on entanglement
swapping. By this protocol \emph{Alice} can securely share a random quantum
key with \emph{Bob}, without transporting any particle.
\end{abstract}

\section{Introduction}

The main goals of cryptography are securely supplying a key to two parties
and detecting eavesdroppers. These goals cannot be achieved classically, but
they can be overcome by quantum cryptography. Since \emph{BB84\ }protocol %
\cite{1},the first quantum cryptography scheme, was published, various
quantum encryption schemes have been proposed, such as the EPR 
scheme \cite{2}\cite{r5a} , the $4+2$ protocol \cite{r5p}, the six-state 
protocol\cite{r5p2}, the Goldenberg /Vaidman scheme\cite{r5p3}, Koashi/Imoto 
scheme \cite{r5p4}, and the recent protocol \cite{long} and so on. But most
of protocol need transport particles.

In this paper,we present a new method in which the communicators use
entanglement swapping\cite{swapping} to distribute the quantum cryptographic
random key and use the correlations of entanglement swapping results to
detect the eavesdropper. There are two merits than the previous protocol in
our scheme, $1.$if the communicators shared enough known entangled pairs
before the key distribution,sender $\left( \text{\emph{Alice}}\right) $ need
not send any particle to receiver $\left( \text{\emph{Bob}}\right) $. $2.$
The efficiency can approach $4$ bit secret communication per $2$ entangled
pairs (in \emph{BB84\ }protocol, it is only $1$ bit per pair particle).

\section{protocol}

Entanglement swapping has been proposed by Zukowski et al.\cite{swapping}
for two entangled pairs in one of the \emph{Bell} states. Entanglement
swapping\ works as follows. Consider two pairs of entangled particles $1$,$2$%
,$3$ and $4$, prepared in \emph{Bell} states respectively $\left| \phi
^{+}\right\rangle _{12}$ and $\left| \psi ^{-}\right\rangle _{34}$. If a 
\emph{Bell} operator measurement is performed on particle $2$ and particles $%
3$, then we have

\begin{eqnarray}
\left| \Psi \right\rangle _{1234} &=&\left| \phi ^{+}\right\rangle
_{12}\otimes \left| \psi ^{-}\right\rangle _{34}  \label{11} \\
&=&\frac{1}{2}\left\{ \left| \phi ^{+}\right\rangle _{23}\left| \psi
^{-}\right\rangle _{14}+\left| \phi ^{-}\right\rangle _{23}\left| \psi
^{+}\right\rangle _{14}-\left| \psi ^{+}\right\rangle _{23}\left| \phi
^{-}\right\rangle _{14}-\left| \psi ^{-}\right\rangle _{23}\left| \phi
^{+}\right\rangle _{14}\right\} .  \notag
\end{eqnarray}

From above equation, we found the four possible result $\left| \phi
^{+}\right\rangle _{23}$,$\left| \phi ^{-}\right\rangle _{23}$,$\left| \psi
^{+}\right\rangle _{23}$ and $\left| \psi ^{-}\right\rangle _{23}$ have the
same probability to occur. In fact ,the outcome of each measurement is
purely random. Suppose that the resulut $\left| \phi ^{+}\right\rangle _{23}$
is obtained, consequebtly the state of the pair $1$ and $4$ after the
measurement is $\left| \psi ^{-}\right\rangle _{14}$. Therefore, the state
of 1 and 4 become s entangled \ although they have never interacted.

Recently entanglement swapping has been used in quantum key distribution\cite%
{rcabello}. Although there are $3$ entangled pairs in protocol\cite{rcabello}%
, yet \emph{Alice} must send particles to \emph{Bob}. Here we propose a
quantum key distribution scheme without particles transported. The proposed
protocol for quantum key distribution is illustrated in Fig. $1$ and it is
described as follows. The parties, \emph{Alice} and \emph{Bob}, initially
share $2n$ \emph{EPR} pairs in known entangled states $\left| \eta
\right\rangle _{AB}$, $\left| \eta \right\rangle _{AB}$ is one of the four 
\emph{Bell} states. \emph{Alice} and \emph{Bob} divide all entangled pairs
into $n$ ordered groups, every group include $2$ entangled pairs. \emph{Alice%
} and \emph{Bob} agree beforehand that $\left| \phi ^{+}\right\rangle $,$%
\left| \psi ^{-}\right\rangle $,$\left| \psi ^{+}\right\rangle $,$\left|
\phi ^{-}\right\rangle $ are encoded as $00$,$01$,$10$,$11$ respectively.
They can perform the series operations as followsing,(\emph{Alice }own
particle $1$ and particle $3$, the particle $2$ and particle $4$ belong to 
\emph{Bob}. )

$1$.\emph{Alice} made a \emph{Bell }operator measurement on particle $1$\
and particle $3$.

For example, we can assumed initially the state of particle $1$ and particle 
$2$ is $\left| \phi ^{+}\right\rangle _{AB}^{12}$, and the state of particle 
$3$ and particle $4$ is $\left| \psi ^{+}\right\rangle _{AB}^{34}$, where

\begin{equation*}
\left| \phi ^{+}\right\rangle _{AB}^{12}=\frac{1}{\sqrt{2}}\left( \left|
00\right\rangle _{AB}^{12}+\left| 11\right\rangle _{AB}^{12}\right) \text{
and}
\end{equation*}

\begin{equation*}
\left| \psi ^{+}\right\rangle _{AB}^{34}=\frac{1}{\sqrt{2}}\left( \left|
01\right\rangle _{AB}^{34}+\left| 10\right\rangle _{AB}^{34}\right) \text{.}
\end{equation*}

\bigskip \emph{Alice} perform \emph{Bell} operator measurement on particle $%
1 $ and particle $3$, and record the measurement.

$2$.\emph{Alice} calculated out the state of particle $2$ and particle $4$.

Suppose that \emph{Alice}'s \emph{Bell} operator measurement is $\left| \psi
^{+}\right\rangle _{AA}^{13}$, she can obtain \emph{Bob}'s \emph{Bell}
operator measured result$\left| \phi ^{+}\right\rangle _{BB}^{24}$ by the
following equation,

\begin{equation}
\left| \Phi \right\rangle _{ABAB}^{1234}=\frac{1}{2}\left\{ \left| \phi
^{+}\right\rangle _{AA}^{13}\left| \psi ^{+}\right\rangle _{BB}^{24}+\left|
\phi ^{-}\right\rangle _{AA}^{13}\left| \psi ^{-}\right\rangle
_{BB}^{24}+\left| \psi ^{+}\right\rangle _{AA}^{13}\left| \phi
^{+}\right\rangle _{BB}^{24}+\left| \psi ^{-}\right\rangle _{AA}^{13}\left|
\phi ^{-}\right\rangle _{BB}^{24}\right\} \text{.}  \label{22}
\end{equation}

\emph{Alice} decode the \emph{Bell} operator measured results,she get $10$
and $00$.

$3$.\emph{Alice} told \emph{Bob} she had made \emph{Bell }operator
measurement on her particles,but did not told him the measured result by
classical channel.

$4$.\emph{Bob} performed a \emph{Bell }operator measurement on particle $2$
and particle $4$ and calculated \emph{Alice}'s \emph{Bell} operator
measurement result.

After he captured \emph{Alice}'s classical information, \emph{Bob} make 
\emph{Bell} operator measurement on particle $2$ and $4$. He get the
measured result $\left| \phi ^{+}\right\rangle _{BB}^{24}$, and he obtain 
\emph{Alice}'s \emph{Bell} operator measurement$\left| \psi
^{+}\right\rangle _{AA}^{13}$ from the eq.($2$). So \emph{Bob} can gain the
bits $10$ and $00$.

they perform above progrcee repeatedly,

$5$.\emph{Bob} chose some results and sent them to \emph{Alice}, then \emph{%
Alice}\ compared\emph{\ Bob}'s results with her corresponding results, she
can find if there is eavesdropper.

\bigskip

The important point is that the efficiency of the quantum key distribution
has been improved in this protocol. From above operations, We know,that 
\emph{Alice} shared $4$ bits with \emph{Bob} by two entangled pairs. This
way largely improves the efficiency of the distribution for cryptographic
key. The efficiency can reach $1$ bit per paticle, it is higher than the
previous protocol, for exampal, in BB84 protocol it is $1$ bit per pair, and
in B92 protocol it is $1$ bit per two pairs.

\section{Security}

How secure is this protocol? Now we discuss the security of our protocol
against three type of attacks.

Type I. General eavesdropping

Suppose that,\emph{Eve},the eavesdropper, can get secretly the classical
information of \emph{Alice} and \emph{Bob} and know all the states of the
quantum channel. If \emph{Alice} and \emph{Bob} do not publish their's \emph{%
Bell} operator measured result, eavesdropper (\emph{Eve}) can not get the
information from the known entangled channels. For example, both the quantum
channels are the same \emph{Bell} states $\left| \phi ^{+}\right\rangle
_{AB}^{12}$ and $\left| \phi ^{+}\right\rangle _{AB}^{34}$. By the
entanglement swapping calculation, we know the state of these entangled
particles is

\begin{equation}
\left| \Phi \right\rangle _{ABAB}^{1234}=\frac{1}{2}\left\{ \left| \phi
^{+}\right\rangle _{AA}^{13}\left| \phi ^{+}\right\rangle _{BB}^{24}+\left|
\phi ^{-}\right\rangle _{AA}^{13}\left| \phi ^{-}\right\rangle
_{BB}^{24}+\left| \psi ^{+}\right\rangle _{AA}^{13}\left| \psi
^{+}\right\rangle _{BB}^{24}+\left| \psi ^{-}\right\rangle _{AA}^{13}\left|
\psi ^{-}\right\rangle _{BB}^{24}\right\} .
\end{equation}

\emph{Eve} have many entangled pairs in state $\left| \phi ^{+}\right\rangle
_{EE}$,she performed Bell operator measurement on her particles $1^{/}$,$%
2^{/}$,$3^{/}$ and $4^{/}$,

\begin{eqnarray}
\left| \Phi ^{/}\right\rangle _{EBEB}^{1^{/}2^{/}3^{/}4^{/}} &=&\frac{1}{2}%
\{\left| \phi ^{+}\right\rangle _{EE}^{1^{/}3^{/}}\left| \phi
^{+}\right\rangle _{BB}^{2^{/}4^{/}}  \label{11111} \\
&&+\left| \phi ^{-}\right\rangle _{EE}^{1^{/}3^{/}}\left| \phi
^{-}\right\rangle _{BB}^{2^{/}4^{/}}+\left| \psi ^{+}\right\rangle
_{EE}^{1^{/}3^{/}}\left| \psi ^{+}\right\rangle _{BB}^{2^{/}4^{/}}+\left|
\psi ^{-}\right\rangle _{EE}^{1^{/}3^{/}}\left| \psi ^{-}\right\rangle
_{BB}^{2^{/}4^{/}}\}.  \notag
\end{eqnarray}

The four \emph{Bell }operator measurements in eq.[$3$] and [$4$] are equally
likely, each occurring with probability $\dfrac{1}{4}$. The probability, 
\emph{Eve's} \emph{Bell }operator measurements on particles $1^{/}$ and $%
3^{/}$ is same as \emph{Alice's Bell }operator measurements on particles $1$
and $3$,is only $\dfrac{1}{4}$. That is to say,the probability, \emph{Eve}
eavesdrop $4n$ bits and is not detected by \emph{Alice} or \emph{Bob}, is
only $\left( \dfrac{1}{4}\right) ^{n}$. When $n$ is large enough, this
probability is $0$ approximately.

Type II. Share the quantum channel.

If the \emph{Eve} were smart enough, she made the state $\left| \phi
^{+}\right\rangle _{ABE}^{\prime }=\frac{1}{\sqrt{2}}\left( \left|
00\right\rangle _{AB}\left| 0\right\rangle _{E}+\left| 11\right\rangle
_{AB}\left| 1\right\rangle _{E}\right) $ instead of $\left| \phi
^{+}\right\rangle _{AB}$, when \emph{Alice} made a \emph{Bell} operator
measurement on particle $1$ and particle $3$, the state became

\begin{eqnarray}
\left| \Psi \right\rangle _{ABE}^{123456} &=&\left| \phi ^{+}\right\rangle
_{ABE}^{\prime }\otimes \left| \phi ^{+}\right\rangle _{ABE}^{\prime }
\label{1111} \\
&=&\dfrac{1}{2\sqrt{2}}\{\left| \phi ^{+}\right\rangle _{AA}^{13}\left(
\left| \phi ^{+}\right\rangle _{BB}^{24}\left| \phi ^{+}\right\rangle
_{EE}^{56}+\left| \phi ^{-}\right\rangle _{BB}^{24}\left| \phi
^{-}\right\rangle _{EE}^{56}\right)  \notag \\
&&+\left| \phi ^{-}\right\rangle _{AA}^{13}\left( \left| \phi
^{+}\right\rangle _{BB}^{24}\left| \phi ^{-}\right\rangle _{EE}^{56}+\left|
\phi ^{-}\right\rangle _{BB}^{24}\left| \phi ^{+}\right\rangle
_{EE}^{56}\right)  \notag \\
&&+\left| \psi ^{+}\right\rangle _{AA}^{13}\left( \left| \psi
^{+}\right\rangle _{BB}^{24}\left| \psi ^{+}\right\rangle _{EE}^{56}+\left|
\psi ^{-}\right\rangle _{BB}^{24}\left| \psi ^{-}\right\rangle
_{EE}^{56}\right)  \notag \\
&&+\left| \psi ^{-}\right\rangle _{AA}^{13}\left( \left| \psi
^{+}\right\rangle _{BB}^{24}\left| \psi ^{-}\right\rangle _{EE}^{56}+\left|
\psi ^{-}\right\rangle _{BB}^{24}\left| \psi ^{+}\right\rangle
_{EE}^{56}\right) \}.  \notag
\end{eqnarray}

Compare eq.[$3$] with eq.[$5$] , we found if there is an eavesdropper, there
is only $\dfrac{1}{2}$ probability that the two results are same. \emph{Bob}
can send some results randomly to Alice; then Alice will be sure that there
is an eavesdropper if she finds different results---they would then abandon
this key. That is, Eve cannot gain any information about the key.

Type III. Replace the quantum channel

If \emph{Eve }shared the entangled pairs $\left| \phi ^{+}\right\rangle
_{AE}^{ij}$ with \emph{Alice} instead of \emph{Bob}, and \emph{Eve }shared
entangled pairs $\left| \phi ^{+}\right\rangle _{EB}^{i^{/}j^{/}}$ with 
\emph{Bob}, before the key supply, both \emph{Alice} and \emph{Bob} would be
unaware of this. \ Then the process of key distribution become\ 

\begin{equation}
\left| \Phi \right\rangle _{AEAE}^{1234}=\frac{1}{2}\left\{ \left| \phi
^{+}\right\rangle _{AA}^{13}\left| \phi ^{+}\right\rangle _{EE}^{24}+\left|
\phi ^{-}\right\rangle _{AA}^{13}\left| \phi ^{-}\right\rangle
_{EE}^{24}+\left| \psi ^{+}\right\rangle _{AA}^{13}\left| \psi
^{+}\right\rangle _{EE}^{24}+\left| \psi ^{-}\right\rangle _{AA}^{13}\left|
\psi ^{-}\right\rangle _{EE}^{24}\right\}  \label{444}
\end{equation}

and

\begin{equation}
\left| \Phi ^{/}\right\rangle _{EBEB}^{1^{/}2^{/}3^{/}4^{/}}=\frac{1}{2}%
\left\{ \left| \phi ^{+}\right\rangle _{EE}^{1^{/}3^{/}}\left| \phi
^{+}\right\rangle _{BB}^{2^{/}4^{/}}+\left| \phi ^{-}\right\rangle
_{EE}^{1^{/}3^{/}}\left| \phi ^{-}\right\rangle _{BB}^{2^{/}4^{/}}+\left|
\psi ^{+}\right\rangle _{EE}^{1^{/}3^{/}}\left| \psi ^{+}\right\rangle
_{BB}^{2^{/}4^{/}}+\left| \psi ^{-}\right\rangle _{EE}^{1^{/}3^{/}}\left|
\psi ^{-}\right\rangle _{BB}^{2^{/}4^{/}}\right\} .  \label{555}
\end{equation}

From Eq.($6$) and Eq.($7$), we know that, the probability, which \emph{Bell }%
operator measurement of particle $1$ and $3$ is same as particle $1^{/}$ and 
$3^{/}$, is only $\dfrac{1}{4}$, when \emph{Bob} send some his measurement
results to \emph{Alice}, \emph{Alice} can find if there is eavesdropper.

So we can say this protocal is secure.

\section{Conclusion}

We have given an effective quantum key distribution protocol,if the two
legitimate partities shared much enough entangled pairs before the random
key distribution, then there is no particle to be transported. The
efficiency can approach 1 bit secret communication per particle.

\begin{equation*}
\FRAME{itbpFU}{4.5904in}{3.2076in}{0in}{\Qcb{Fig.$1$ Random quantum key
distribution protocol}}{\Qlb{Picture 1}}{random.jpg}{\special{language
"Scientific Word";type "GRAPHIC";maintain-aspect-ratio TRUE;display
"USEDEF";valid_file "F";width 4.5904in;height 3.2076in;depth
0in;original-width 6.4273in;original-height 4.4789in;cropleft "0";croptop
"1";cropright "1";cropbottom "0";filename 'random.jpg';file-properties
"XNPEU";}}
\end{equation*}

\end{document}